\documentclass[a4paper,11pt]{article}
\usepackage[dvips]{graphicx}
\usepackage{amssymb}
\usepackage{amsmath}

\usepackage{cite}

\begin{document}

\title{Study of Antigravity in an F(R) Model and in Brans-Dicke Theory with Cosmological Constant}
\author{
V.K. Oikonomou\,\thanks{voiko@physics.auth.gr}\\
Department of Theoretical Physics, Aristotle University of Thessaloniki,\\
54124 Thessaloniki, Greece\\\\
N. Karagiannakis\,\thanks{nikar@auth.gr}\\
Polytechnic School, Aristotle University of Thessaloniki,\\
54124 Thessaloniki, Greece
} \maketitle

\begin{abstract}
We study antigravity, that is having an effective gravitational constant with a negative sign,
in scalar-tensor theories originating from $F(R)$-theory and in a Brans-Dicke model with
cosmological constant. For the $F(R)$ theory case, we obtain the antigravity
scalar-tensor theory in the Jordan frame by using a variant of the Lagrange multipliers method and
we numerically study the time dependent effective gravitational constant. As we shall demonstrate by
using a specific $F(R)$ model, although there is no antigravity in the initial model, it might occur
or not in the scalar-tensor counterpart, mainly depending on the parameter that characterizes
antigravity. Similar results hold true in the Brans-Dicke model. 
\end{abstract}

\section*{Introduction}

During the last two decades our perception about the universe has changed drastically owing to the
discovered late time acceleration that our universe has. Particularly, it can be thought as one of
the most striking astrophysical observations with another striking observation being the
verification of the inflating period of our universe. Actually, moving from time zero to present
time, inflation came first, with the late time acceleration occurring at present epoch. One of the
greater challenges in cosmology is to model this late time acceleration in a self consistent way.
According to the new Planck telescope observational data for the present epoch, the universe is
consistently described by the $\mathrm{\Lambda}\mathrm{CDM}$ model, according to which the universe
is nearly spatially flat, and consists of ordinary matter ($\sim4.9\%$), cold dark matter ($\sim
26.8\%$) and dark energy ($\sim 68.3\%$). The dark energy is actually responsible for late time
acceleration and current research on the field is mostly focused on this issue.

One of the most promising and theoretically appealing descriptions of dark energy and late time acceleration issues, is provided by the 
$F(R)$ modified theories of gravity and related modifications. For important review articles and papers on the vast issue of $F(R)$ theories, the reader is referred to \cite{reviews1,reviews2,reviews3,reviews4,reviews5,reviews6,reviews7,importantpapers1,importantpapers2,importantpapers3,importantpapers4,importantpapers5,importantpapers6,importantpapers7,importantpapers8,importantpapers9,importantpapers10,importantpapers11,importantpapers12,importantpapers13} and references therein. For some alternative theories to modified gravity that model dark energy, see \cite{bambaalt,capo,capo1,peebles,faraonquin,tsujiintjd}. The most appealing characteristic of modified gravity theories theories is that, what is actually changed is not the left hand side of the Einstein equations, but the right hand side. Late time acceleration then, requires a negative $w$ fluid, which can be consistently incorporated in the energy momentum tensor of these theories. This feature naturally appears in $
 F(R)$ theories and also late time acceleration solutions of the Friedmann-Robertson-Walker equations, naturally occur in these theoretical frameworks \cite{reviews1,reviews2,reviews3,reviews4,reviews5,reviews6,reviews7,importantpapers1,importantpapers2,
importantpapers3,importantpapers4,importantpapers5,importantpapers6,importantpapers7,
importantpapers8,importantpapers9,importantpapers10,importantpapers11,importantpapers12,
importantpapers13,sergeinojirimodel}. In addition, inflation, the first accelerating period of our
universe, is also consistently described by some $F(R)$ theories, rendering the latter a very
elegant and economic description of nature at large scales, where General Relativity fails to
describe phenomena consistently. Particularly, the possibility to theoretically describe in a
consistent and elegant way, early-time inflation and late-time acceleration in $F(R)$ gravity was
explicitly demonstrated in the Nojiri-Odintsov model \cite{sergeinojirimodel}. For studies on
specific solutions in several strong curved backgrounds see
\cite{solutions1,solutions2,solutions3,solutions4,solutions5,solutions6}. Remarkable possibilities, like modified gravitational theories with non-minimal curvature-matter coupling, were given in \cite{bertolami1,bertolami2,bertolami3,bertolami4} and references therein. 

In principle, every consistent generalizations of general relativity inevitably has to be confronted with the successes of general relativity. Since general relativity is a successful description of nature in strong gravitational environments, there exist a large number of constraints need to be satisfied, in order an $F(R)$ modified gravity theory can be considered as viable. The constraints to be satisfied are mainly imposed from local tests of general relativity, for example from planetary and star formation tests and moreover from various cosmological bounds. In addition, since each $F(R)$ theory has a Jordan frame scalar-tensor gravitational theory counterpart, with $\omega$ zero and a potential, the scalarons of this counterpart theory must be classical, in order to ensure quantum-mechanical stability (see \cite{reviews1,reviews2,reviews3,reviews4,reviews5,reviews6,reviews7}). 

In theories of modified gravity a longstanding debatable theoretical problem exists, related to
Jordan and Einstein frames \cite{fujii,faraoni}, since the physics coming out from the two
frames can be quite different in principle. In view of this, we shall focus on the physics of Jordan
frame and demonstrate that it is possible to have antigravity.
\cite{antigravity1,antigravity2,antigravity3,antigravityfrodi} For the possibility of
antigravity regimes in scalar-tensor theories consult \cite{antigravity1,antigravity2,antigravity3}
and for antigravity in $F(R)$ theories see \cite{antigravityfrodi}. In this paper we shall study
antigravity regimes coming from $F(R)$ theories and from Brans-Dicke theories in the Jordan frame.
In reference to the $F(R)$ theories, we shall find the Jordan frame antigravity scalar-tensor
counterpart, using a modified method of the Lagrange multipliers, as we shall see in the following
sections. The interesting feature about these theories is that, although
the $F(R)$ theory has no antigravity, the resulting Jordan frame scalar-tensor theory may or may not
have antigravity. We exemplify this by numerically working out an example. In the case of
Brans-Dicke antigravity, we introduce by hand an antigravity term and numerically solve the
cosmological equations and as we shall demonstrate, similar results hold true, that is, antigravity
may exist or not, depending on the parameters of the theory.

This paper is organized as follows: In section 1 we briefly recall the essentials of $F(R)$
theories, in section 2 we get to the core of the paper and introduce a modification of the Lagrange
multipliers method in order to get antigravity from $F(R)$ theories. Accordingly we apply the
technique to one quite known $F(R)$ model and present the result of our analysis. The study of
antigravity is performed in section 3 and the conclusions follow in the end of the paper.

\section{General Features of $F(R)$ Dark Energy Models in the Jordan Frame}

In this section in order to maintain the article self-contained, we briefly review the main features of $F(R)$ gravity theories in the Jordan frame in the theoretical framework of the metric formalism. For an important stream of review papers and articles see \cite{reviews1,reviews2,reviews3,reviews4,reviews5,reviews6,reviews7,importantpapers1,importantpapers2,importantpapers3,importantpapers4,importantpapers5,importantpapers6,importantpapers7,importantpapers8,importantpapers9,importantpapers10,importantpapers11,importantpapers12,importantpapers13}  and references therein. 

The geometrical background of the manifolds used here is pseudo-Riemannian and is described locally
by a Lorentz metric (the FRW metric in our case), in addition to a torsion-less, symmetric, and
metric compatible affine connection, the so-called Levi-Civita connection. In such a geometric
background, the Christoffel symbols are
\begin{equation}\label{christofell}
\Gamma_{\mu \nu }^k=\frac{1}{2}g^{k\lambda }(\partial_{\mu }g_{\lambda \nu}+\partial_{\nu
}g_{\lambda \mu}-\partial_{\lambda }g_{\mu \nu})
\end{equation} 
and the Ricci scalar becomes
\begin{equation}\label{ricciscalar}
R=g^{\mu \nu }(\partial_{\lambda }\Gamma_{\mu \nu }^{\lambda}-\partial_{\nu }\Gamma_{\mu \rho
}^{\rho}-\Gamma_{\sigma \nu }^{\sigma}\Gamma_{\mu \lambda }^{\sigma}+\Gamma_{\mu \rho }^{\rho}g^{\mu
\nu}\Gamma_{\mu \nu }^{\sigma}).
\end{equation}
The $F(R)$ theories of modified gravity are described by a modification of the Einstein-Hilbert action, with the four dimensional action being equal to:
\begin{equation}\label{action}
\mathcal{S}=\frac{1}{2\kappa^2}\int \mathrm{d}^4x\sqrt{-g}F(R)+S_m(g_{\mu \nu},\Psi_m),
\end{equation}
where $\kappa^2=8\pi G$ and $S_m$ the matter action containing the matter fields $\Psi_m$. For simplicity in this section it shall be assumed that the form of the $F(R)$ theory that will be used is $F(R)=R+f(R)$ and in addition the metric formalism framework shall be used. Varying the action (\ref{action}) with respect to the metric $g_{\mu \nu}$, we get the following equations of motion:
\begin{equation}\label{eqnmotion}
F'(R)R_{\mu \nu}(g)-\frac{1}{2}F(R)g_{\mu \nu}-\nabla_{\mu}\nabla_{\nu}F'(R)+g_{\mu \nu}\square F'(R)=\kappa^2T_{\mu \nu}.
\end{equation} 
In the above equation, $F'(R)=\partial F(R)/\partial R$ and also $T_{\mu \nu}$ is the energy momentum tensor. 

What is the most striking feature of the $F(R)$ modified gravity theories is that, what actually changes in reference to the usual Einstein-Hilbert gravity equations, is the right hand side of the Einstein equations and not the left, which remains the same. Indeed, the equations of motion (\ref{eqnmotion}) can be cast in the following form:
\begin{align}\label{modifiedeinsteineqns}
R_{\mu \nu}-\frac{1}{2}Rg_{\mu \nu}=\frac{\kappa^2}{F'(R)}\Big{(}T_{\mu \nu}+\frac{1}{\kappa}\Big{[}\frac{F(R)-RF'(R)}{2}g_{\mu \nu}+\nabla_{\mu}\nabla_{\nu}F'(R)-g_{\mu \nu}\square F'(R)\Big{]}\Big{)}.
\end{align}
Therefore we get an additional contribution for the energy momentum tensor, coming from the term:
\begin{equation}\label{newenrgymom}
T^{eff}_{\mu \nu}=\frac{1}{\kappa}\Big{[}\frac{F(R)-RF'(R)}{2}g_{\mu
\nu}+\nabla_{\mu}\nabla_{\nu}F'(R)-g_{\mu \nu}\square F'(R)\Big{]}.
\end{equation}
It is this term that actually models the dark energy in $F(R)$ theories of modified gravity. Taking the trace of equation (\ref{eqnmotion}) we straightforwardly obtain the following equation:
\begin{equation}\label{traceeqn}
3\square F'(R)+R F'(R)-2F(R)=\kappa^2 T,
\end{equation}
where $T$ stands for the trace of the energy momentum tensor $T=g^{\mu \nu}T_{\mu \nu}=-\rho+3P$,
and, additionally, $\rho$ and $P$ stand for the matter energy density and pressure respectively.

There exists another degree of freedom in $F(R)$ theories, as can be easily seen by observing
equation (\ref{traceeqn}). This degree of freedom is actually a scalar degree of freedom, called 
scalaron, described by the function $F'(R)$, with equation (\ref{traceeqn}) being the equation of
motion of this scalar field. In a flat Friedmann-Lemaitre-Robertson-Walker spacetime, the Ricci
scalar is equal to:
\begin{equation}\label{ricciscal}
R=6(2H^2+\dot{H}),
\end{equation}
with $H$ being the Hubble parameter and the ``dot'' indicating differentiation with respect to time.
The cosmological equations of motion are given by the following set of equations:
\begin{subequations}\label{flrw}
  \begin{align}
     3F'(R)H^2&=\kappa^2(\rho_m+\rho_r)+\frac{(F'(R)R-F(R))}{2}-3H\dot{F}'(R),\text{ and} \\
    -2F'(R)\dot{H}&=\kappa^2(p_m+4/3\rho_r)+F\ddot{F}'(R)-H\dot{F}'(R),
  \end{align}
\end{subequations}
with $\rho_r$ and $\rho_m$ standing for the radiation and matter energy density respectively. Thereby, the total effective energy density and pressure of matter and geometry are \cite{reviews1,reviews2,reviews3,reviews4,reviews5,reviews6,reviews7}:
\begin{subequations}\label{densitypressure}
  \begin{align}
    \rho_{eff}&=\frac{1}{F'(R)}\Big{[}\rho_m+\frac{1}{\kappa^2}
      \Big{(}F'(R)R-F(R)-6H\dot{F}'(R)\Big{)}\Big{]},\text{ and} \\ 
     p_{eff}&=\frac{1}{F'(R)}\Big{[}p_m+\frac{1}{\kappa^2}
      \Big{(}-F'(R)R+F(R)+4H\dot{F}'(R)+2\ddot{F}'(R)\Big{)}\Big{]},
  \end{align} 
\end{subequations}
where $\rho_m,P_m$ denote the total matter energy density and matter pressure respectively.

\section{Antigravity in $F(R)$ Models}

The possibility of antigravity sectors in $F(R)$ theories was firstly pointed out in
\cite{antigravityfrodi} and also in various scalar-tensor models in references
\cite{antigravity1,antigravity2,antigravity3}. In most cases, a passing from antigravity to a
gravity regime always occurs, with a singularity existing at the transition between these two
different gravitational regimes. At the transition, the effective gravitational constant and also
several invariants of the geometry, such as the Weyl invariant, become singular quantities
\cite{antigravityfrodi,antigravity1,antigravity2,antigravity3}. In the present article, we are
interested in studying the time dependence of the effective gravitational constant and see how this
behaves for both an $F(R)$ theory related antigravity scalar-tensor model and an
antigravity version of the Brans-Dicke model with cosmological constant. In reference to $F(R)$
theories, we shall explicitly demonstrate in the next subsection how to find the antigravity
scalar-tensor theory in the Jordan frame. By doing so, we will have at hand an a\-nti\-gra\-vi\-ty
scalar-tensor theory with a potential term and we shall explicitly find how the scalar field, along
with the gravitational constant and the energy density, behaves for various values of the model
dependent and cosmological variables. Then we study the Brans-Dicke model
in which we shall make a by hand modification in order to render it an antigravity model. As we
shall see, in both cases, there exist several gravity-antigravity regimes, depending on the values
of the model dependent and cosmological variables. Moreover, for the $F(R)$-model, although the
model per-se has no antigravity, the corresponding scalar-tensor model gives rise to antigravity
regimes. However, there exist values of the variables for which the model describes gravity regimes.
In the following subsections we shall study in detail these models.

\subsection{A General Way to Obtain Antigravity Scalar-Tensor Models from $F(R)$ models}

It is a quite well known fact that scalar-tensor theories are equivalent to $F(R)$ theories. In the
literature one starts from an $F(R)$ theory and ends up to a non-minimally coupled scalar-tensor
theory and more specifically to a Brans-Dicke theory with $\omega_{BD}$ equal to zero. This is
practically the Lagrange multipliers method (see
\cite{reviews1,reviews2,reviews3,reviews4,reviews5,reviews6,reviews7} and particularly Odintsov and
Nojiri (2007) and Felice and Tsujikawa (2010)). 

In this paper we shall use a variant but quite similar method to obtain an antigravity
scalar-tensor theory starting from a given $F(R)$ theory. Consider the general $F(R)$ theory with
matter, which is described by the action
\begin{equation}\label{actionlagmult}
\mathcal{S}=\int \mathrm{d}^4x\sqrt{-g}F(R)+S_m(g_{\mu \nu},\Psi_m).
\end{equation}
Introducing an auxiliary field $\chi$, which acts as a Lagrange multiplier, the action
(\ref{actionlagmult}) becomes
\begin{equation}\label{actionlagmult1}
\mathcal{S}=\int \mathrm{d}^4x\sqrt{-g}\Big{(}F(\chi)+F_{,\chi}(\chi)(R-\chi)\Big{)}+S_m(g_{\mu \nu},\Psi_m)
\end{equation}
with $F_{,\chi}(\chi)$ being the first derivative of the function $F(\chi)$ with respect to $\chi$. By varying the action (\ref{actionlagmult1}) with respect to $\chi$ we obtain:
\begin{equation}\label{fgegdffdbkf}
F_{,\chi \chi}(\chi)(R-\chi)=0.
\end{equation}
Given that $F_{,\chi \chi}(\chi)\neq 0$, which is actually true for most viable $F(R)$ theories, we
may conclude that $R=\chi$. Hence, the action (\ref{actionlagmult1}) actually recovers the initial
$F(R)$-gravity action (\ref{actionlagmult}). We define,
\begin{equation}\label{newdef}
\varphi-\mathcal{B}=F_{,\chi}(\chi)
\end{equation}
and the action of equation (\ref{actionlagmult1}) is expressed as a function of the field $\varphi$
in the following way:
\begin{equation}\label{actionlagmult123}
\mathcal{S}=\int \mathrm{d}^4x\sqrt{-g}\Big{[}\big{(}\varphi-\mathcal{B}\big{)} R-U(\varphi
)\Big{]}+S_m(g_{\mu \nu},\Psi_m).
\end{equation}
Comparing the non-minimal coupling term $\left(\varphi-\mathcal{B}\right)R$ to the
corresponding term $\frac{1}{16\pi G}R$ of the standard Einstein-Hilbert action, we
get the relation for the effective gravitational constant
\begin{equation}
  G_{eff}=\frac{1}{16\pi\left(\varphi-\mathcal{B}\right)}.
\end{equation}
It is easy to see that if $\varphi(t)-\mathcal{B}=F_{,\chi}(\chi)<0$ there emerges antigravity. The potential term  $U(\varphi )$ is equal to
\begin{equation}\label{potentialbdfr}
U(\varphi )=\chi (\varphi)\big{(}\varphi-\mathcal{B}\big{)} -F\big{(}\chi (\varphi )\big{)},
\end{equation}
where the function $\chi (\varphi)$ is directly obtained by solving the algebraic equation
(\ref{newdef}) with respect to $\chi$, so that $\chi$ is an explicit function of $\varphi$.
Therefore as result, starting from an $F(R)$ theory and using the technique we just presented,
one obtains Jordan frame antigravity scalar-tensor theories.

\subsection{The Model $F(R)=R-R^{-p}$ with $p$ a Positive Integer}

As an application of the method we just presented, let us use a viable $F(R)$ model a modified version of which is quite frequently used in $F(R)$ cosmology \cite{reviews1,reviews2,reviews3,reviews4,reviews5,reviews6,reviews7}. The model has the following form as a function of the curvature scalar $R$:
\begin{equation}\label{frmod1scq}
F(R)=R-R^{-p},
\end{equation} 
with $p$ being some positive integer number. This form of the $F(R)$ function ensures that the first
derivative of the $F(R)$ function with respect to $R$ is positive definite for $R\geq R_D$, with
$R_D$ being the final de-Sitter attractor solution of the theory, that is
 \begin{equation}\label{condantigrfr}
 \frac{\mathrm{d}F(R)}{\mathrm{d}R}>0.
\end{equation}
The condition (\ref{condantigrfr}) assures that no antigravity occurs for the $F(R)$ model \cite{reviews1,reviews2,reviews3,reviews4,reviews5,reviews6,reviews7}. However, as we shall demonstrate, antigravity might occur in the Jordan frame scalar-tensor model. 
\begin{figure}[ht]
\begin{minipage}{17pc}
\includegraphics[width=14pc]{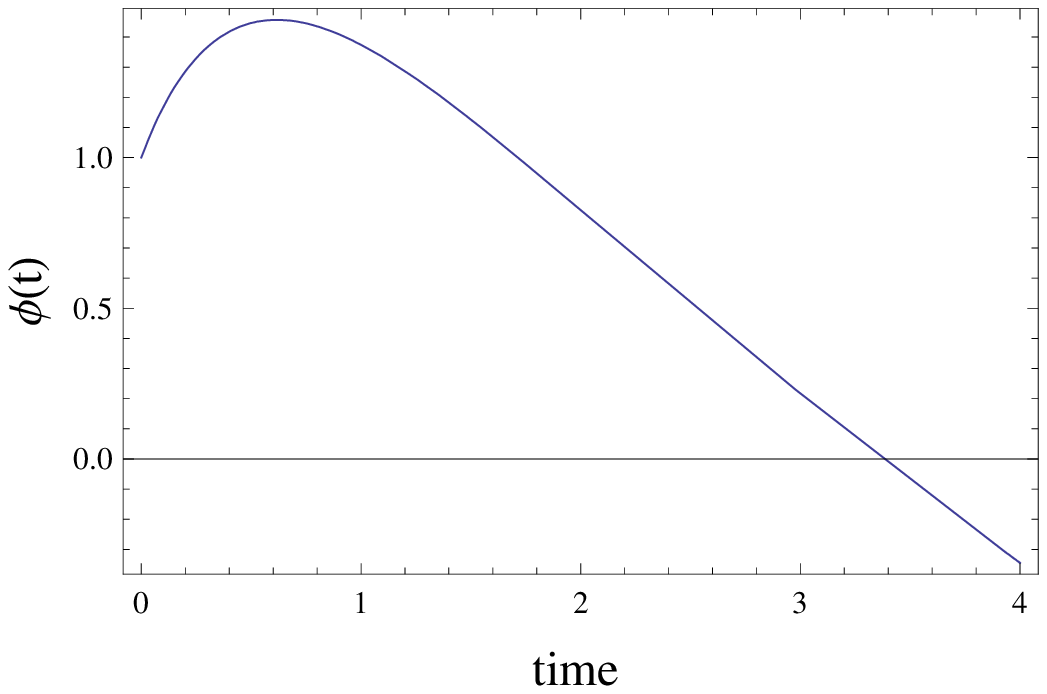}
\end{minipage}
\begin{minipage}{17pc}
\includegraphics[width=15pc]{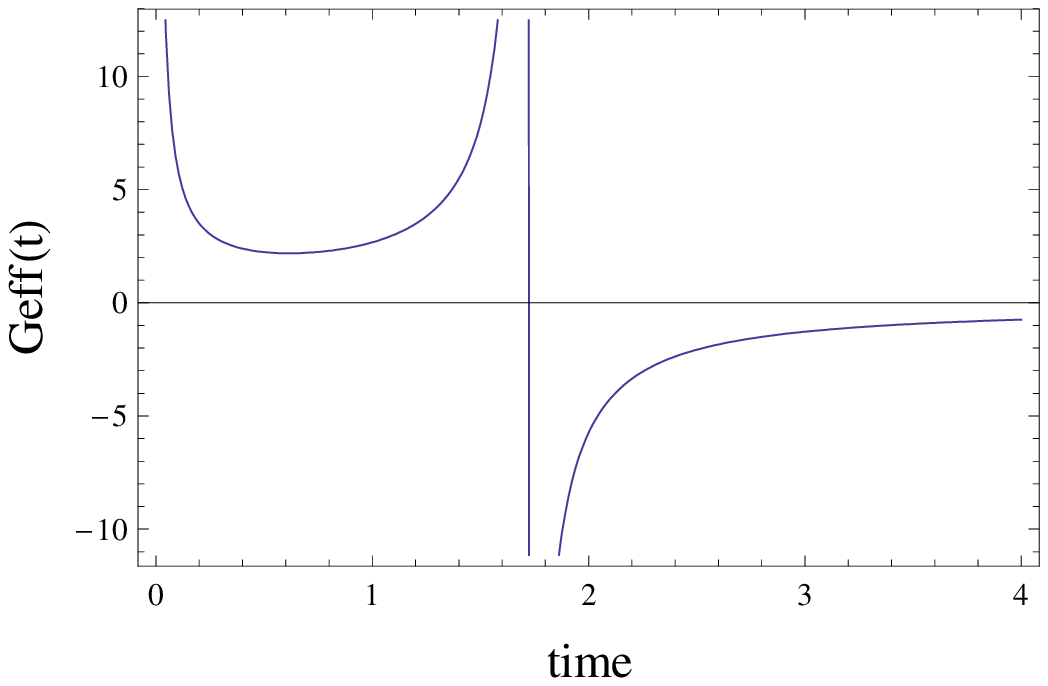}
\end{minipage}
\begin{minipage}{17pc}
\includegraphics[width=14pc]{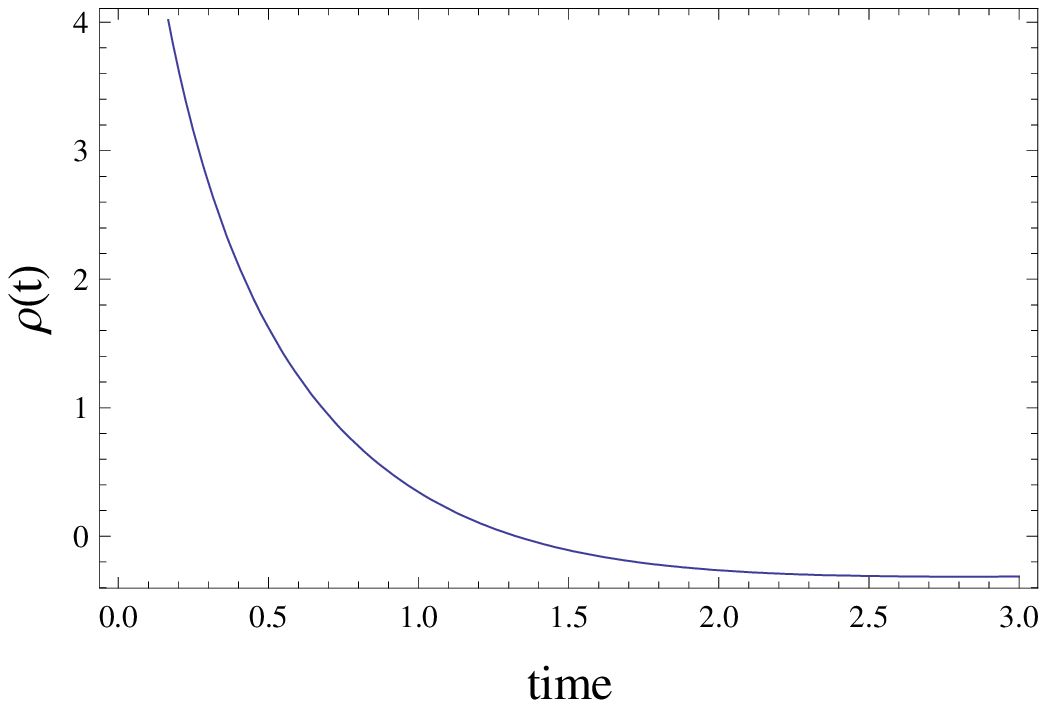}\hspace{2pc}%
\end{minipage}
\begin{minipage}[b]{17pc}
\caption{\label{enasc4} $F(R)$ model: Time dependence of the scalar field $\varphi (t)$ (left), the
effective gravitational constant $G_{eff}(t)$ (right) and the matter energy density $\rho (t)$
(bottom), for $w=\frac{1}{3}$, $p=3$, $\mathcal{B}=1$}
\end{minipage}
\end{figure}
The action corresponding to the $F(R)$ action (\ref{frmod1scq}) is the following,
\begin{equation}\label{actionlagmultsc1}
\mathcal{S}=\int \mathrm{d}^4x\sqrt{-g}\big{(}R-R^{-p}\big{)}+S_m(g_{\mu \nu},\Psi_m).
\end{equation}
Using the Lagrange multipliers method we introduced in the previous section, we obtain the corresponding scalar-tensor antigravity theory, with the Jordan frame action being equal to:
\begin{equation}\label{actionlagmult123sc1}
\mathcal{S}=\int \mathrm{d}^4x\sqrt{-g}\Big{[}\big{(}\varphi-\mathcal{B}\big{)} R-U_{F(R)}(\varphi
)\Big{]}+S_m(g_{\mu \nu},\Psi_m).
\end{equation}
The potential $U_{F(R)}(\varphi )$ for the present $F(R)$ model is equal to
\begin{equation}\label{potentialbdfrsc1}
U_{F(R)}(\varphi
)=\Big{(}\frac{p}{\varphi-\mathcal{B}-1}\Big{)}^{\frac{1}{p+1}}(\varphi-\mathcal{B})-\Big{(}\frac{p}
{\varphi-\mathcal{B}-1}\Big{)}^{\frac{1}{p+1}}+\Big{(}\frac{p}{\varphi-\mathcal{B}-1}\Big{)}^{-\frac
{p}{p+1}}.
\end{equation}
Having action (\ref{actionlagmult123sc1}) at hand, along with potential term
(\ref{potentialbdfrsc1}), we can study the antigravity scalar-tensor model in a straightforward way.
By varying action (\ref{actionlagmult123sc1}) with respect to the metric and the
scalar field, we get the Einstein equations that describe the cosmic evolution of the antigravity
$F(R)$-related scalar-tensor model. Assuming a flat FRW metric of the form
\begin{equation}\label{metricformfrw}
\mathrm{d}s^2=-\mathrm{d}t^2+a^2(t)\sum_i\mathrm{d}x_i^2
\end{equation}
the cosmological equations are equal to
\begin{figure}[ht]
\begin{minipage}{18pc}
\includegraphics[width=14pc]{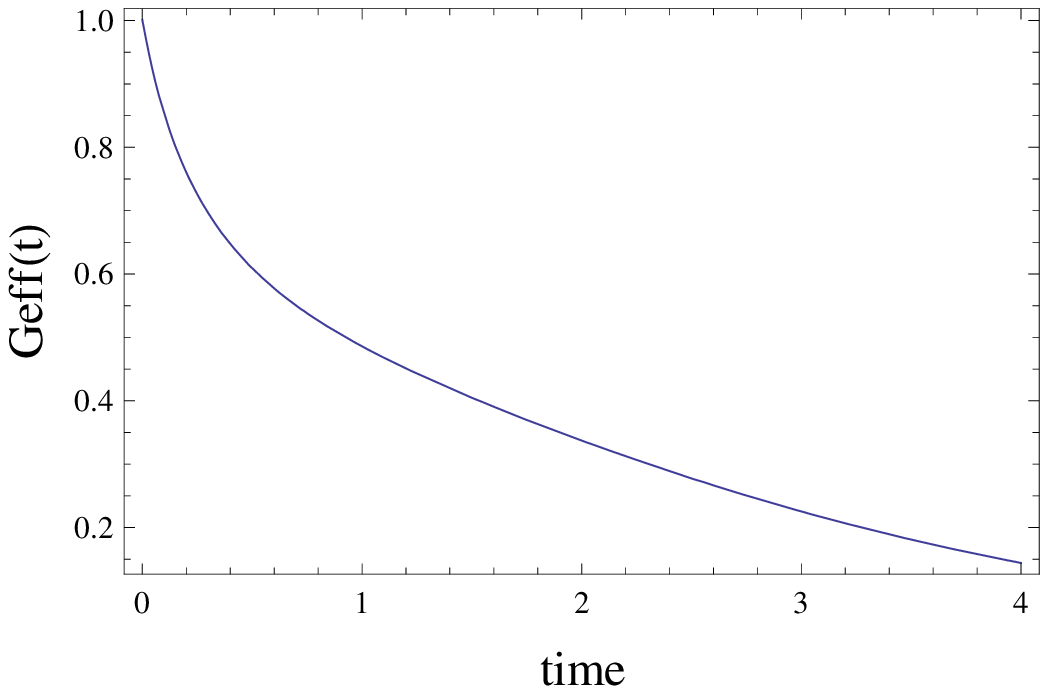}
\end{minipage}
\begin{minipage}{17pc}\hspace{2pc}%
\includegraphics[width=15pc]{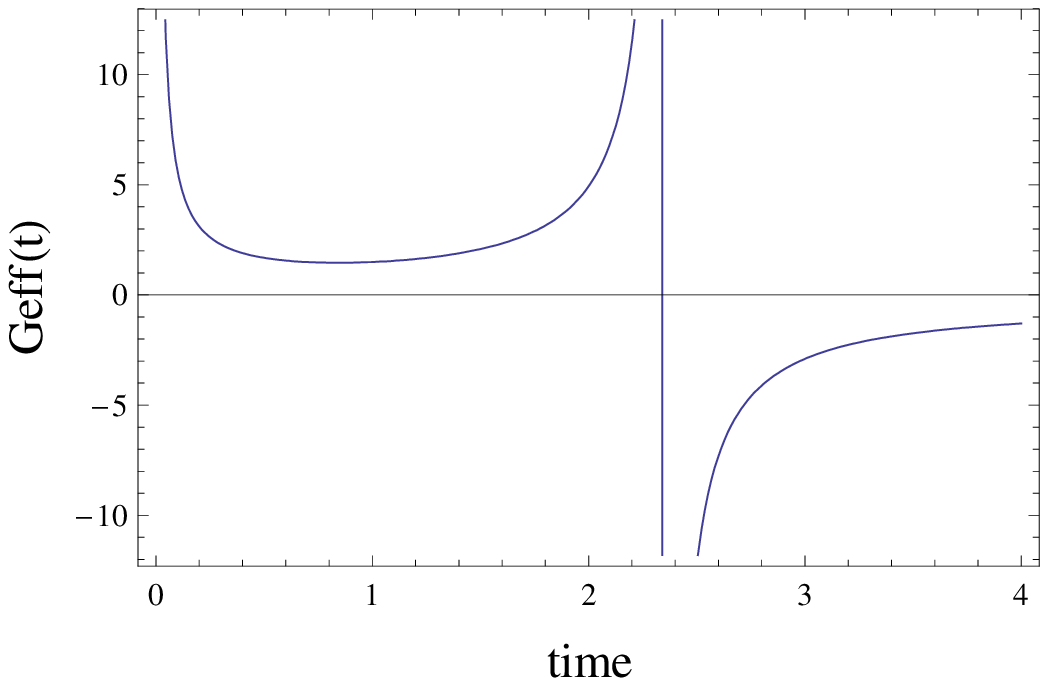}
\end{minipage}
\caption{\label{enasc3}$F(R)$ model: The effective gravitational constant
$G_{eff}(t)$ as a function of time, for non-relativistic matter $w=0$, with $\mathcal{B}=0.001$,
$p=2$ (left) and $\mathcal{B}=1$, $p=2$ (right)}
\end{figure}
\begin{subequations}\label{feqnssc}
  \begin{align}
    3(\varphi-\mathcal{B} )H^2&=\rho+U_{F(R)}(\varphi )-3H\dot{\varphi} \\ 
    -2 (\varphi-\mathcal{B} )\dot{H}&=\rho+P+\ddot{\varphi}-H\dot{\varphi} \\
    -R+2\frac{\mathrm{d}U_{F(R)} (\varphi )}{\mathrm{d}\varphi }&=0, 
  \end{align}
\end{subequations}
where $\dot{\varphi }$ denotes differentiation of the scalar field function $\varphi (t)$, with respect to the time variable t:
\begin{equation}\label{hfgf}
\dot{\varphi}=\frac{\mathrm{d}\varphi }{\mathrm{d}t }.
\end{equation}
In addition $P=w\rho$ and also the continuity equation for matter, stemming from $T^{\mu \nu
}_{;\mu}=0$,  holds true:
\begin{equation}\label{energconsv}
\dot{\rho }+3H(1+w)\rho=0.
\end{equation}
From equation (\ref{actionlagmult123sc1}), it easily follows that the effective gravitational constant of the Jordan frame scalar-tensor theory, is equal to:
\begin{equation}\label{effgravcnstfr1}
G_{eff}(t)=\frac{1}{16\pi (\varphi (t)-\mathcal{B})}.
\end{equation}
We numerically solved the cosmological equations (\ref{feqnssc}) and in Figures 1 and 2 we
present the results which we will now analyze in detail. As a general comment let us note that,
depending on the value of the antigravity parameter $\mathcal{B}$, the Jordan frame scalar-tensor
theory may or may not have antigravity. Therefore, although we started with an $F(R)$ theory with no
antigravity solutions, the Jordan frame counterpart exhibits antigravity for some values of the
parameter $\mathcal{B}$. In order for the time dependent functions $\varphi$, $\rho$ and $G_{eff}$
to vary smoothly, we chose the initial conditions to be 
\begin{equation}
  \rho(1)=1,\quad \varphi(1)=1,\quad \dot{\varphi}(1)=0, \quad \text{and}\quad t\times H(1)\sim1,
\end{equation}
which are similar to those used in reference \cite{fujii2} (check also \cite{fujii}). We also
performed the following rescaling for time:
\begin{equation}
  t=1 \rightarrow 10^{-46} \mathrm{sec},
\end{equation}
in favor of the simplicity of the plots. The above initial conditions and time scaling are used
for all the plots in this article. The results obtained by the numerical analysis
are qualitatively robust towards the change of the initial conditions, meaning that the
only thing that changes is not the whole phenomenon, but the exact time point when the
singularity occurs; in all cases the transition singularity occurs long before the beginning of
inflation at $t=10^{10}\rightarrow 10^{-36} \mathrm{sec}$. In Figure 1 we provide plots of the
scalar field $\varphi (t)$,
the energy
density $\rho (t)$, and the effective gravitational constant $G_{eff}(t)$ as a function of the time
$t$, with the time axis properly rescaled. We have chosen the numerical values to be
$w=\frac{1}{3}$, $p=3$, and $\mathcal{B}=1$, that is in a radiation dominated universe. The
same behavior however is observed for $\mathcal{B}=1$ and different values for $w$. Therefore, we
observe that the parameter $\mathcal{B}$ critically affects the antigravity behavior. In the
present case, the occurring antigravity can be seen in the right part of Figure 1; as can be seen,
there appears a gravity dominated period for $0<t<1.7$ and after the singularity at
$t=1.7$ antigravity occurs. In Figure 2, we present the time dependence of the effective
gravitational constant $G_{eff}(t)$, for two different values of $\mathcal{B}$, namely
$\mathcal{B}=0.001$ (left) and $\mathcal{B}=1$ (right). We assumed a universe filled with
non-relativistic matter, that is $w=0$, and also $p=2$. As we can see in this case, for
$\mathcal{B}=0.001$ there is no antigravity and conversely for $\mathcal{B}=1$ there is. This is the
expected behavior of the Jordan frame theory, since as $\mathcal{B}$ increases, the possibility that
the term $(\varphi (t) -\mathcal{B})$ becomes negative increases, depending, of course, on the
initial conditions and on the other parameters' values. 

The model we studied in this section is similar to the one studied in \cite{antigravityfrodi}, in which case the antigravity scalar-tensor model was the following:
\begin{align}\label{generaleqnbdantigrav}
S=&\int \mathrm{d}x^4\sqrt{-g}\Big{[}\frac{1-\varphi^2}{12} R-\frac{1}{2}g^{\mu
\nu}\partial_{\mu}\varphi \partial_{\nu}\varphi -J(\varphi )\Big{]}.
\end{align}
The corresponding $F(R)$ gravity action,following the technique presented in
\cite{antigravityfrodi} is easily found to be 
\begin{align}\label{generaleqn4}
S=&\int \mathrm{d}x^4\sqrt{-g}F(R),
\end{align}
where $F(R)$ stands for:
\begin{equation}\label{frequation1}
F(R)=\frac{e^{\eta (\varphi (R))}}{12}\Big{(}1-\varphi^2(R)\Big{)}R-e^{2\eta (\varphi (R))}J(\varphi
(R)).
\end{equation}
Moreover, the real function $\eta(\varphi )$ satisfies
\begin{equation}\label{cnstr}
(1+2\varphi^2)\eta'(\varphi)^2-4\eta'(\varphi )-4=0
\end{equation} 
and, as a result, the kinetic term of the scalar field vanishes. This antigravity model clearly
provides us with regimes governed by a negative gravitational constant for some values of the scalar
field $\varphi $, clearly indicating a highly non-smooth, big
crunch-big bang transition in the theoretical context of \cite{antigravityfrodi}.

Before we close this section, we discuss an important issue. Reasonably, it can be argued that since
the effective gravitational constant $G_{eff}(t)$ diverges at some time, this could imply some sort
of instability of the $F(R)$ theory. Indeed, this is true to some extend. Actually, the singularity
of the gravitational constant is a spacetime one, since spacetime geometric invariants
like the Kretschmann scalar $R_{abcd}R^{abcd}$ seriously diverge. In a mathematical
context, this singularity is also a naked Cauchy horizon, not ``dressed'' by some event horizon,
which, in turn, would imply the loss of predictability and also signal a spacetime singularity.
Therefore, it is better if these singularities occur in the very early universe. As for the issue of
stability of the initial $F(R)$ theory, this is an involved question, since the quantum mechanical
stability of the $F(R)$ theory is examined in the Einstein frame and not in the Jordan frame
\cite{reviews1}. In the case of an occurring singularity, the Einstein frame is not
consistently defined, since this singularity also introduces another singularity in the scalar field
redefinition necessary for the definition of the canonical transformation in the Einstein frame (see
the book of Faraoni for more details on this \cite{faraoni}). A very thorough analysis of the
stability of a, similar to ours, scalar-tensor model was studied in reference \cite{vernov} (see
equation (1) of \cite{vernov}), in which case the model can exhibit antigravity if the non-minimal
coupling term becomes negative. The model in \cite{vernov} can be identical to our Brans-Dicke model
if the potential is zero and the non-minimal coupling contains terms of the order of $\sim \varphi$.


\section{Antigravity in Brans-Dicke Models}

As we saw in the previous section, even though we started from an $F(R)$ theory with no antigravity,
the antigravity Jordan frame action may or may not have antigravity solutions. In this section, we
shall study a minor modification of the Brans-Dicke model with cosmological constant. The
antigravity term will be introduced by hand and will be of the form $\big{(}\varphi
-\mathcal{B}\big{)}R$, with $\mathcal{B}$ being the extra term introduced by hand. 
\begin{figure}[ht]
\begin{minipage}{18pc}
\includegraphics[width=14pc]{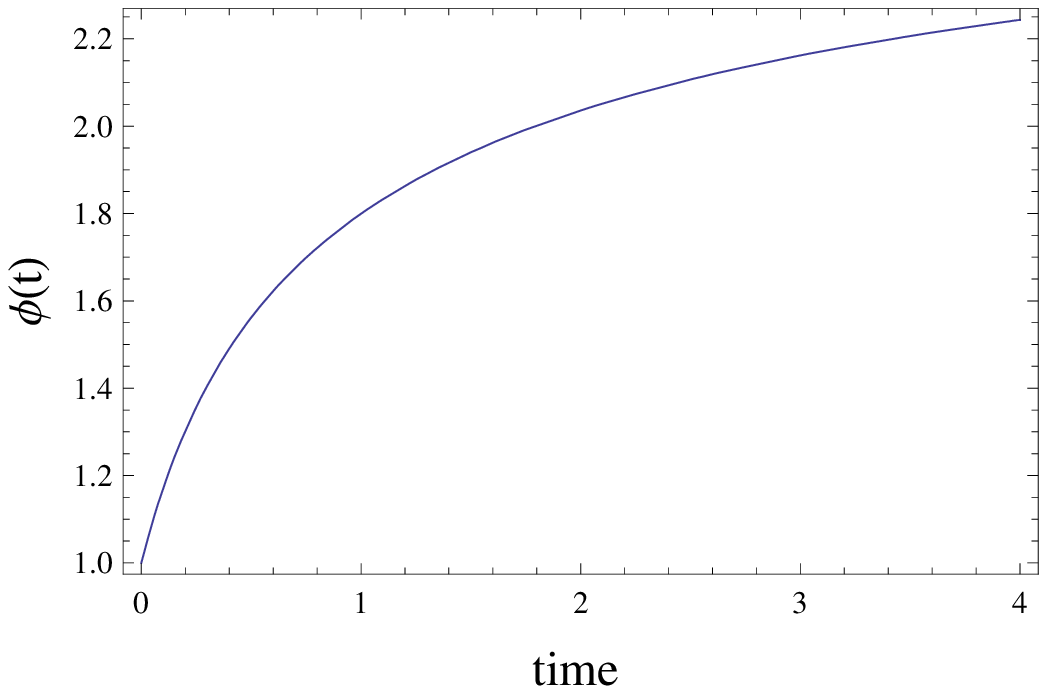}
\end{minipage}
\begin{minipage}{17pc}\hspace{2pc}%
\includegraphics[width=15pc]{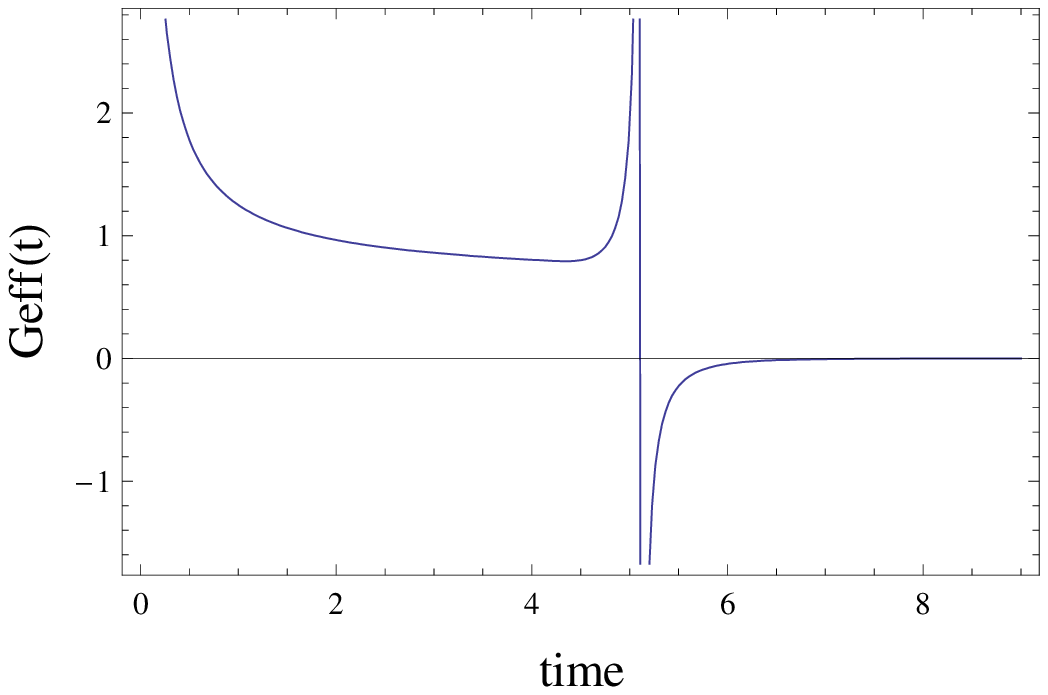}
\end{minipage}
\includegraphics[width=14pc]{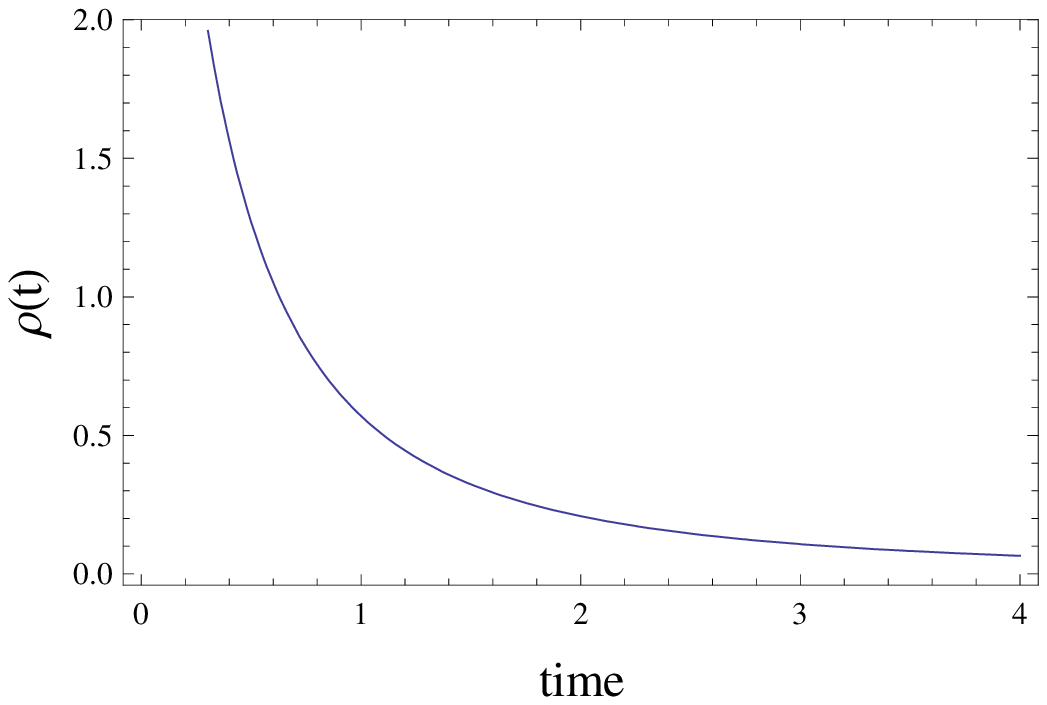}\hspace{2pc}%
\begin{minipage}[b]{14pc}\caption{\label{enasc2} Brans-Dicke Model with Cosmological constant: Time dependence of the scalar field $\varphi (t)$ (left), the effective gravitational constant $G_{eff}(t)$ (right) and the matter energy density $\rho (t)$ (bottom), for $w=\frac{1}{3}$, $\mathcal{B}=1$, and the cosmological constant $\Lambda =10^{-49}$.}
\end{minipage}
\end{figure}
The general action in the Jordan frame that describes a general Brans-Dicke model with cosmological
constant, potential $U(\varphi )$, and matter is:
\begin{align}\label{generaleqnbdsc1}
S=&\int \mathrm{d}x^4\sqrt{-g}\Big{[}\frac{1}{2}\varphi
\big{(}R-2\Lambda\big{)}-\frac{\omega_{BD}}{\varphi }g^{\mu \nu}\partial_{\mu}\varphi
\partial_{\nu}\varphi-U(\varphi ) \Big{]} +\int \mathrm{d}x^4\sqrt{-g}L_{matter}.
\end{align}
In the following, we shall assume that initially, the scalar potential $U(\varphi )$ is zero and
also that the cosmological constant is positive and has the value $\Lambda =10^{-49}\mathrm{GeV}^4$.
\begin{figure}[ht]
\begin{minipage}{18pc}
\includegraphics[width=14pc]{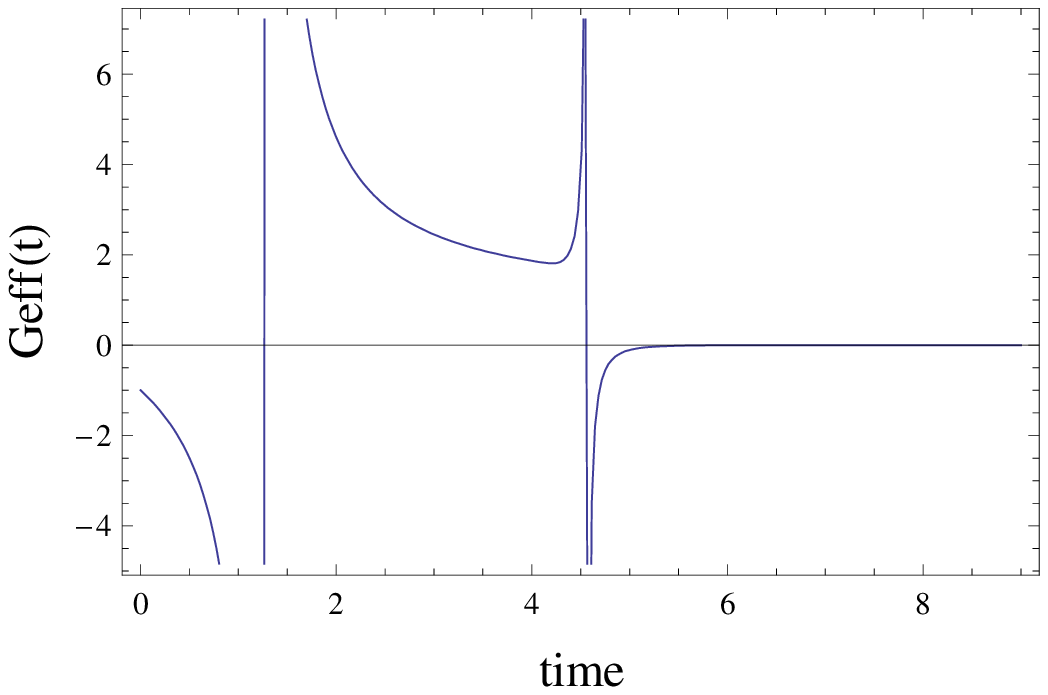}
\end{minipage}
\begin{minipage}{17pc}\hspace{2pc}%
\includegraphics[width=15pc]{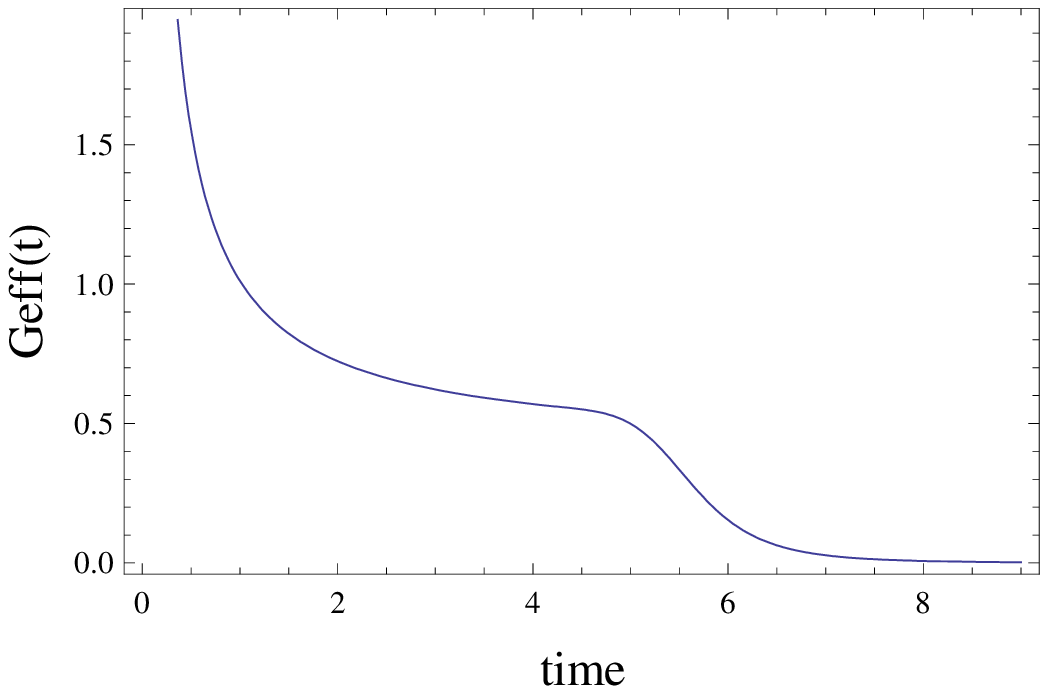}
\end{minipage}
\caption{\label{enasc1} Brans-Dicke Model with Cosmological constant: The effective gravitational
constant $G_{eff}(t)$ as a function of time, for non-relativistic matter $w=0$, with $\Lambda
=10^{-49}$, and $\mathcal{B}=2$ (left) and $\mathcal{B}=1$ (right)}
\end{figure}
The antigravity model we shall study is obtained from the original Brand-Dicke model with cosmological constant (\ref{generaleqnbdsc1}) if we modify by hand the action in the following way:
\begin{align}\label{generaleqnbdsc1sc11}
S=&\int \mathrm{d}x^4\sqrt{-g}\Big{[}\frac{1}{2}\big{(}\varphi
-\mathcal{B}\big{)}R-\frac{\omega_{BD}}{\varphi }g^{\mu \nu}\partial_{\mu}\varphi
\partial_{\nu}\varphi-\varphi \lambda \Big{]} +\int \mathrm{d}x^4\sqrt{-g}L_{matter}.
\end{align}
The term $\varphi \lambda$ acts as a potential term and hence we have at hand an antigravity Brans-Dicke model with potential $U_{BD}(\varphi )=\varphi \Lambda$. 
By varying \ref{generaleqnbdsc1sc11} with respect to the metric and the scalar
field, we obtain the Einstein equations describing the cosmological evolution of the antigravity
Brans-Dicke model, which for a flat FRW metric are equal to
\begin{subequations}\label{feqnsscsbd}
  \begin{align}
 3\big{(}\varphi-\mathcal{B}\big{)}H^2&=\rho^{(m)}+\frac{\omega_{BD}}{2}\Big{(}\dot{\varphi
}\Big{)}^2+\varphi \Lambda-3H\dot{\varphi}, \\ 
-2 \big{(}\varphi-\mathcal{B}\big{)}\dot{H}&=\rho^{(m)}+P^{(m)}+\omega_{BD}\Big{(}\dot{\varphi
}\Big{)}^2+\ddot{\varphi}-H\dot{\varphi},\text{ and} \\ 
\ddot{\varphi }+3H\dot{\varphi }+\frac{1}{2\omega_{BD}}\Big{[}-R+2\Lambda\Big{]}&=0.
  \end{align}
\end{subequations}
In the following, we shall take $\omega_{BD}=1/2$. As in the previous case, the effective
gravitational constant varies with time in the Jordan frame model and its value is given by
\begin{equation}\label{effgravcnbdst}
G_{eff}(t)=\frac{1}{16\pi (\varphi (t)-\mathcal{B})}.
\end{equation}
We have solved numerically the cosmological equations (\ref{feqnsscsbd}) and as a general remark let
us note that the model has both gravity and antigravity solutions, depending on the values of the
parameters and specifically on the value of the antigravity parameter $\mathcal{B}$. In Figures 3
and 4 we have presented the results of our numerical analysis for various parameter values and we
now discuss in detail. In Figure 1 appears the time dependence of the scalar field $\varphi (t)$,
the energy density $\rho (t)$, and the effective gravitational constant $G_{eff}(t)$, where again we
have properly rescaled the time axis. The numerical values we used in Figure 3 are $w=\frac{1}{3}$,
$\mathcal{B}=1$, and $\Lambda =10^{-49}$. Changing
the value of $w$ does not drastically affect the solutions, which crucially depend on the value of
the antigravity parameter $\mathcal{B}$. As can be seen from the time dependence of the effective
gravitational constant $G_{eff}(t)$ in Figure 3, antigravity occurs along with a singularity between
the transition from gravity to antigravity. This latter feature is quite common in antigravity
models (see for example \cite{antigravity1,antigravity2,antigravity3,antigravityfrodi}).
Accordingly, in Figure 4 we have provided the plots of the effective gravitational constant as a
function of time, for $w=0$, $\Lambda =10^{-49}$, and $\mathcal{B}=2(1)$ for the left (right) plot.
Obviously, for $\mathcal{B}=2$ (left) a complex antigravity pattern occurs, while for
$\mathcal{B}=1$ (right) there is no antigravity at all. This result validates our observation
that antigravity crucially depends on the values of the $\mathcal{B}$.

\section{A Brief Discussion}

Before closing this section, we discuss a last issue of some importance. It is generally known that a general $F(R)$ theory with the method of Lagrange multipliers can be transformed to a Brans-Dicke theory with $\omega_{BD}=0$ and non zero potential. Indeed, it is easy to see this and we demonstrate it shortly. Consider a general $F(R)$ theory described by the following action:
\begin{equation}\label{actionlagmultjogrev}
\mathcal{S}=\int \mathrm{d}^4x\sqrt{-g}F(R)+S_m(g_{\mu \nu},\Psi_m)
\end{equation}
We introduce an auxiliary field $\chi$, which actually is the Lagrange multiplier. Using this field, the action (\ref{actionlagmultjogrev}) becomes:
\begin{equation}\label{actionlagmult1jogrev}
\mathcal{S}=\int \mathrm{d}^4x\sqrt{-g}\Big{(}F(\chi)+F_{,\chi}(\chi)(R-\chi)\Big{)}+S_m(g_{\mu
\nu},\Psi_m),
\end{equation}
with $F_{,\chi}(\chi)$ being the first derivative of the function $F(\chi)$ with respect to $\chi$.
Varying the action (\ref{actionlagmult1jogrev}), with respect to the auxiliary field $\chi$, we get,
\begin{equation}\label{fgegdffdbkfjogrev}
F_{,\chi \chi}(\chi)(R-\chi)=0.
\end{equation}
Recalling that $F_{,\chi \chi}(\chi)\neq 0$, which actually holds true for most viable $F(R)$
theories, we get  $R=\chi$. Therefore, the action (\ref{actionlagmult1jogrev}) recovers the initial
$F(R)$-gravity action (\ref{actionlagmult1jogrev}). If we define,
\begin{equation}\label{newdefjogrev}
\varphi=F_{,\chi}(\chi),
\end{equation}
then the action appearing in equation (\ref{actionlagmult1jogrev}) becomes actually a function
of the field $\varphi$, as can be seen below:
\begin{equation}\label{actionlagmult123jogrev}
\mathcal{S}=\int \mathrm{d}^4x\sqrt{-g}\Big{[}\varphi R-U(\varphi )\Big{]}+S_m(g_{\mu \nu},\Psi_m)
\end{equation}
The scalar potential term  $U(\varphi )$ is equal to the following expression:
\begin{equation}\label{potentialbdfrjogrev}
U(\varphi )=\chi (\varphi)\varphi -F\big{(}\chi (\varphi )\big{)}.
\end{equation}
By solving the algebraic equation (\ref{newdefjogrev}) with respect to $\chi $, will actually give
us in closed form the function $\chi (\varphi)$ (at least in most cases), as a function of $\varphi$. Therefore, it is a
straightforward way to obtain a Brans-Dicke theory with $\omega_{BD}$ zero and non-zero potential by
starting from a general $F(R)$ theory. A question naturally springs to mind, that is, whether it is
possible to have any sort of coincidence between $F(R)$ gravity and Brans-Dicke with a non-zero
potential and zero $\omega_{BD}$ and the answer is actually yes, but only when the potential of the
Brans-Dicke is exactly the one of equation (\ref{potentialbdfrjogrev}). Now, one has to be cautious,
however, because this coincidence is ``one way'' only, meaning that if we start with the Brans-Dicke
theory with $\omega_{BD}=0$ and we try to find the
corresponding $F(R)$ theory by using a conformal transformation, then we may end up with a
different $F(R)$ theory, which we denote for example $f(R)$. This requires a much deeper study
that extends beyond the purpose of this article and we defer this interesting issue to a near future
work. However, the reader is referred to the method in four dimensions used by the authors in
\cite{antigravityfrodi}. There, it can be seen that, when starting from a general scalar-tensor
theory, we end up with a certain class of $F(R)$ theories determined by a constraint which the
scalar field has to obey. It is not obvious, however, that starting from a Brans-Dicke theory with
$\omega_{BD}=0$ and non-zero potential, we will end up to the original $F(R)$ theory we started
with. We hope to answer this issue in a future publication.



\section{Conclusions} 
 
In this paper we studied antigravity in scalar-tensor theories originating from $F(R)$ theories and
also antigravity in the Brans-Dicke model with cosmological constant. In the case of the $F(R)$
theories we used a variant of the Lagrange multipliers method leading to antigravity scalar-tensor
model in the Jordan frame, with $\omega =0$ and a scalar potential. We applied the technique
and studied numerically the time-dependence of the gravitational constant. As we exemplified,
although the initial $F(R)$ model has no antigravity, guaranteed by the condition $F'(R)>0$, the
scalar-tensor Jordan frame counterpart may or may not have antigravity. This latter feature strongly
depends on the parameters of the theory and particularly on the antigravity parameter $\mathcal{B}$.
In the case of the Brans-Dicke model with cosmological constant, we studied a by hand introduced
antigravity modification of the model in the Jordan frame. The numerical analysis of the
cosmological equations showed that the model exhibits antigravity depending on the numerical values
of the parameters and particularly on the $\mathcal{B}$ antigravity parameter, like in the $F(R)$
model case. In both cases, there exist regimes in the cosmic evolution in which either gravity or
antigravity prevails and when going from antigravity to gravity and vice versa a singularity occurs,
like in most antigravity contexts \cite{antigravity1,antigravity2,antigravity3,antigravityfrodi}. It
worths searching theoretical constructions in which such a singularity is avoided. This would
probably require some sort of singular conformal transformations between frames, or some singularity
of the Lagrangian, a task we hope to address in the near future.

Finally, it is worth discussing the results and also the cosmological implications of our results.
The main goal of this article was to demonstrate all possible cases in which antigravity might
appear in modified theories of gravity. As we explicitly demonstrated, in the case of $F(R)$
theories, although the initial Jordan frame $F(R)$ theory had no antigravity (recall the condition
$F'(R)>0$ which actually guarantees this), antigravity might show up when the Jordan frame
equivalent theory is considered, modified in the way we explicitly showed in the text. This is one
of the new and notable results of this article. In the case of Brans-Dicke model, introducing by
hand a term that causes antigravity, then antigravity might or not appear in the resulting theory.
The latter depends strongly on the value of the antigravity parameter $\mathcal{B}$. In principle,
antigravity is a generally unwanted feature in modified theories of gravity and thus it can be
considered less harmful if it occurs in the very early universe, prior to inflation. Indeed, this
is exactly what happens in all the cases we explicitly demonstrated in the text. However,
antigravity is rather difficult to detect experimentally, unless there exists some mechanism of
creation of a primordial black hole during the antigravity regime that could retain some
information in terms of some sort of gravitational memory \cite{barrow}. The evaporation of this
black hole could reveal the value of the gravitational constant at the time it was
created. A well posed question may be to ask how such a compact gravitational object could be created
in an antigravity regime. The answer to this could be that antimatter
behaves somehow different in antigravity regimes, so it could probably play a prominent
role in such a scenario. However, we have to admit that this is just a speculation, since after
antigravity occurs, the universe experiences a gravitational regime with a spacetime singularity at
the moment of transition. We cannot imagine how a compact gravitational object (if any) could react
under such severe conditions.


\begin{thebibliography}{99}


 


\bibitem{reviews1} S. Nojiri, S. D. Odintsov, Int. J. Geom. Meth. Mod.Phys. 4 (2007) 115

\bibitem{reviews2} A. Felice, Shinji Tsujikawa, Living Rev.Rel. 13 (2010) 3


\bibitem{reviews3} T. P. Sotiriou, V. Faraoni, Rev.Mod.Phys. 82 (2010) 451

\bibitem{reviews4} T. P. Sotiriou, V. Faraoni, Rev.Mod.Phys. 82 (2010) 451


\bibitem{reviews5} S. Nojiri, S. D. Odintsov,  Phys.Rept. 505 (2011) 59

\bibitem{reviews6} K. Bamba, S. Capozziello, S. Nojiri, S. D. Odintsov, Astrophys. Space Sci. 342, 155 (2012)


\bibitem{reviews7}S. Capozziello, M. De Laurentis, Phys.Rept. 509 (2011) 167


\bibitem{importantpapers1} S. Capozziello, S. Nojiri, S.D. Odintsov, A. Troisi, Phys.Lett. B639 (2006) 135

\bibitem{importantpapers2} S. Nojiri, S. D. Odintsov, Gen.Rel.Grav. 36 (2004) 1765


\bibitem{importantpapers3} S. Nojiri, S. D. Odintsov, Phys.Rev. D74 (2006) 086005


\bibitem{importantpapers4} S. Tsujikawa, Phys.Rev. D77 (2008) 023507


\bibitem{importantpapers5} S. Nojiri, S. D. Odintsov, Phys.Lett. B657 (2007) 238


\bibitem{importantpapers6} A. A. Starobinsky, JETP Lett. 86 (2007) 157


\bibitem{importantpapers7} S. M. Carroll, V. Duvvuri, M. Trodden, M. S. Turner, Phys.Rev. D70 (2004) 043528


\bibitem{importantpapers8} O. Bertolami, R. Rosenfeld, Int.J.Mod.Phys. A23 (2008) 4817


\bibitem{importantpapers9} A. Capolupo, S. Capozziello, G. Vitiello, Int.J.Mod.Phys. A23 (2008) 4979


\bibitem{importantpapers10}   P. K.S. Dunsby, E. Elizalde, R. Goswami, S. Odintsov, D. S. Gomez, Phys.Rev. D82 (2010) 023519


\bibitem{importantpapers11}  E.I. Guendelman, A.B. Kaganovich, Int.J.Mod.Phys. A21 (2006) 4373


\bibitem{importantpapers12}  G. Cognola, E. Elizalde, S. Nojiri, S.D. Odintsov, L. Sebastiani, S. Zerbini, Phys.Rev. D77 (2008) 046009


\bibitem{importantpapers13}  S.K. Srivastava, Int.J.Mod.Phys. A22 (2007) 1123


\bibitem{bambaalt} K. Bamba, S. Capozziello, S. Nojiri, S. D. Odintsov, Astrophys.Space Sci. 342 (2012) 155


\bibitem{capo} S. Capozziello, V.F. Cardone, S. Carloni, A. Troisi, Int.J.Mod.Phys. D12 (2003) 1969

 \bibitem{capo1} S. Capozziello, Int.J.Mod.Phys. D11 (2002) 483

\bibitem{peebles} P.J.E. Peebles, Bharat Ratra, Rev.Mod.Phys. 75 (2003) 559

\bibitem{faraonquin} V. Faraoni, Int.J.Mod.Phys. D11 (2002) 471

\bibitem{tsujiintjd} Edmund J. Copeland, M. Sami, Shinji Tsujikawa, Int.J.Mod.Phys. D15 (2006) 1753


\bibitem{sergeinojirimodel} S. Nojiri, S. D. Odintsov, Phys.Rev. D68 (2003) 123512


\bibitem{solutions1} J.P. Morais Graca, V.B. Bezerra, Mod.Phys.Lett. A27 (2012) 1250178

\bibitem{solutions2} M. Sharif, S. Arif, Mod.Phys.Lett. A27 (2012) 1250138


\bibitem{solutions3} S. Asgari, R. Saffari, Gen.Rel.Grav. 44 (2012) 737


\bibitem{solutions4} K. A. Bronnikov, M.V. Skvortsova, A.A. Starobinsky, Grav.Cosmol. 16 (2010) 216


\bibitem{solutions5}  E.V. Arbuzova, A.D. Dolgov, Phys. Lett. B 700 (2011) 289


\bibitem{solutions6} Chung-Chi Lee, Chao-Qiang Geng, L. Yang, Prog. Theor. Phys. 128 (2012) 415


\bibitem{bertolami1}Harko, T., Lobo, F.S.N., Nojiri, S. and Odintsov, S.D., Phys. Rev. D 84(2011)024020

\bibitem{bertolami2}Bertolami, O., Boehmer, C.G. Harko, T. and Lobo, F.S.N., Phys. Rev. D 75 (2007) 104016

\bibitem{bertolami3}Haghani, Z., Harko, T., Lobo, F.S.N., Sepangi, H.R. and Shahidi, S., Phys. Rev. D 88 (2013) 044023

\bibitem{bertolami4}Sharif, M. and Zubair, M.: Astrophys. Space Sci. 349(2014)457


\bibitem{faraoni} Valerio Faraoni, ''Cosmology in Scalar-Tensor Gravity'', Kluwer Academic Publishers, 2004, Netherlands

\bibitem{fujii} Yasunori Fujii, Kei-Ichi Maeda, ''The Scalar-Tensor Theory of Gravitation'', Cambridge University Press, 2004, Cambridge UK
 
\bibitem{antigravity1} P. Caputa, S. S. Haque, J. Olson and B. Underwood, Class. Quant. Grav. 30, 195013 (2013)

\bibitem{antigravity2} I. Bars, S. H. Chen, P. J. Steinhardt, N. Turok, Phys. Lett. B 715, 278 (2012)

\bibitem{antigravity3} J. J. M. Carrasco, W. Chemissany, R. Kallosh, arXiv:1311.3671

\bibitem{antigravityfrodi} K. Bamba, Shin'ichi Nojiri, S. D. Odintsov, Diego Saez-Gomez, Phys. Lett. B 730 (2014) 136

\bibitem{fujii2} Y. Fujii, Prog. Theor. Phys. 99 (1998) 599

\bibitem{vernov} M. A. Skugoreva, A. V. Toporensky, S. Yu. Vernov, arXiv:1404.6226

\bibitem{barrow} J. D. Barrow, Phys.Rev. D46 (1992) R3227-R3230

\end{thebibliography}
\end{document}